\documentclass[prb,superscriptaddress,aps,onecolumn,10pt,showkeys]{revtex4-1}
\usepackage{graphics,graphicx,epsfig,color,latexsym,float,amsmath,bm}
\usepackage[colorlinks,citecolor=blue,linkcolor=blue,urlcolor=blue]{hyperref}
\usepackage[version=4]{mhchem}
\usepackage{wrapfig}
\usepackage{tikz}
\usetikzlibrary{arrows,chains,decorations.pathreplacing,decorations.markings}
\usepackage{amsfonts,amssymb}

\usepackage{ulem}
\usepackage{lipsum}
\usepackage{placeins}
	
\renewcommand{\Re}{\mathop{\mathrm{Re}}}
\renewcommand{\Im}{\mathop{\mathrm{Im}}}

\begin{document}
\title{Time-space bi-fractional drift-diffusion   equation for anomalous electrochemical transport}

\author{Anis Allagui}
\affiliation{Department of Sustainable and Renewable Energy Engineering, University of Sharjah, Sharjah, P.O. Box 27272,
United Arab Emirates}
\affiliation{Department of Mechanical and Materials Engineering, Florida International University, Miami, FL33174, United States}

\author{Georgii Paradezhenko}
\affiliation{Skolkovo Institute of Science and Technology, Moscow 121205, Russia}
\author{Anastasiia Pervishko}
\affiliation{Skolkovo Institute of Science and Technology, Moscow 121205, Russia}
\author{Dmitry Yudin}
\affiliation{Skolkovo Institute of Science and Technology, Moscow 121205, Russia}

\author{Hachemi Benaoum} 
%\email{hbenaoum@sharjah.ac.ae}
\affiliation{Dept. of Applied Physics and Astronomy, 
University of Sharjah, PO Box 27272, Sharjah, United Arab Emirates }

\date{\today}

\begin{abstract}

The Debye-Falkenhagen differential equation is commonly used as a mean-field macroscopic model for describing electrochemical ionic drift and diffusion in dilute binary electrolytes when subjected to a suddenly applied potential smaller than the thermal voltage. However, the ionic transport in most electrochemical systems, such as electrochemical capacitors, permeation through membranes, biosensors and capacitive desalination, the electrolytic medium is interfaced with porous, disordered, and fractal materials which makes the modeling of electrodiffusive transport with the simple planar electrode theory limited. Here we study a possible generalization of the traditional drift-diffusion equation of Debye and Falkenhagen by incorporating both fractional time and space derivatives for the charge density. The nonlocal (global) fractional time derivative takes into account the past dynamics of the variable such as charge trapping effects and thus subdiffusive transport, while the fractional space derivative allows to simulate superdiffusive transport.

\end{abstract}

\maketitle

\section{Introduction}

For the design and fabrication of most electrochemical  devices and systems (e.g. batteries \cite{galuppini2023nonlinear}, fuel cells \cite{tanner1997effect}, electrochemical supercapacitors \cite{hasyim2017prediction, janssen2018transient}, capacitive deionization for water desalination  \cite{biesheuvel2010nonlinear}) the use of spatially-heterogeneous electrodes with porous  structures  is omnipresent. The same is for protein channels or cell membranes which allow  permeation of ions from one electrolytic solution to another \cite{bolintineanu2009poisson}. The movement of charged species in free solutions interfaced with porous structures is known to be a multi-scale process  that involves on the one hand ionic currents over millimetres in length in electroneutral reservoirs and in micrometer-sized macropores, to form,  on the other hand, nanometer-sized electric double layers   in the electrodes’ pores \cite{kant2017theory}. The pores of can be of different sizes and shapes, and can  be enlarged, partially obstructed  or even completely blocked in the course of time.  Therefore, there is a growing interest in developing advanced theories of  porous electrodes in order to better understanding  the electric double layer   phenomena, ion transport and charge storage in   complex   electrochemical systems.

%\subsection{The drift-diffusion equation}

The classical mean-field modelling  of electrodiffusive transport in electrochemistry is done via the well-studied mathematical framework of Poisson–Nernst–Planck (PNP) \cite{, mirzadeh2014conservative}. 
The PNP  model has been also successfully applied to the description of ionic currents in protein channels of biological membranes \cite{jasielec2021electrodiffusion}.  
It is basically a set of coupled nonlinear  equations  with  partial derivatives in time and space of integer-order that capture the dynamics of the electric potential and ionic densities. For the simple case of a   blocking electrochemical system (without Faradaic processes and without fluid flow) with a dilute, symmetric binary electrolyte of  constant material properties that is, the valences of ions $z^+=z^-=z$, diffusivities $D^+ = D^-=D$ and constant dielectric permittivity, independent of time or space, the dimensionless PNP model in the single coordinate $x$ perpendicular to the electrode or membrane (homogeneous in planes perpendicular to the $x$-axis) is constituted of the Nernst–Planck equations for mass conservation  \cite{SB15, schmuck2011modeling}:
\begin{equation}
 {\partial_{\tilde{t}} \tilde{c}^{\pm}} = \partial_{\tilde{x}} \left(   {\partial_{\tilde{x}} {\tilde{c}}^{\pm}} \pm {\tilde{c}}^{\pm} {\partial_{\tilde{x}} \tilde{\phi}} \right)
 \label{eq9}
\end{equation}
with the Poisson's equation: 
\begin{equation}
- \eta^2 \partial^2_{\tilde{x}} \tilde{\phi} =   \tilde{c}^{+} -  \tilde{c}^{-} \label{eq10}
\end{equation}
Here the reduced variables are: 
$\tilde{x}={x}/{l}$  ($l$ is a reference length scale), 
$\tilde{t}={t}/t_D$ ($t_D=l^2/D$ is a reference time scale), 
$\tilde{c}^{\pm}=c^{\pm}/\bar{c}$  (concentrations of positively and negatively charged ions with $\bar{c}$   a reference concentration), 
$\tilde{\phi}=z e \phi/(k_B T)$ ($\phi$ is the electrostatic potential),  
$\tilde{E}=z e l E/(k_B T) = - {\partial_{\tilde{x}} \tilde{\phi}}$ (reduced electric field), 
 $\eta = \lambda_D/l$ with $\lambda_D = \sqrt{\epsilon_f k_B T / ( \bar{c} z^2 e^2)}$. The constants $k_B$  and $e$ are  the Boltzmann constant and the elemental charge, respectively, $\epsilon_f$  is the dielectric permittivity of the solvent  and $T$ is the thermodynamic temperature (both assumed to be a constant).  
 In principle, this problem given by Eqs.\;\ref{eq9} and\;\ref{eq10} retains well enough the essential features of electrodiffusion dynamics \cite{SB15}. 
 
If we further consider the Debye-Falkenhagen linearization  (i.e. system subjected to a suddenly applied potential smaller than the thermal voltage, thus producing small variation of the bulk density of ions with respect to the one in thermodynamic equilibrium \cite{freire2006electrical}),  the PNP model given by Eqs.\;\ref{eq9} and\;\ref{eq10} reduces to the single diffusion-drift equation for the reduced ionic charge density $\tilde{\rho}=\rho/(ze\bar{c})$ (the difference between cationic and anionic concentrations) \cite{bazant2004diffuse}:
\begin{equation}
{\partial_{\tilde{t}} \tilde{\rho}} =  {\partial^2_{\tilde{x}} \tilde{\rho}}  -  ({2\tilde{c}_0}/{\eta^2}) \tilde{ \rho}     
\label{eq6}
\end{equation}
where   ${2\tilde{c}_0} = \tilde{c}^+ + \tilde{c}^-$ is a constant. 
Eq.\;\ref{eq6} can also be rewritten in the form:
\begin{equation}
{\partial_{\tilde{\tau}} \tilde{\rho}} =  {\partial^2_{\tilde{z}} \tilde{\rho}}  -  \tilde{ \rho}     
\label{eq8n} 
\end{equation}
where $\tilde{\tau} =  \tilde{t} ({2\tilde{c}_0}/{\eta^2})$ and $\tilde{z}=  \tilde{x} \sqrt{({2\tilde{c}_0}/{\eta^2})}$. For ease of notation, we shall drop the tildes and replace $\tau$ by $t$ and $z$ by $x$,  such as Eq.\;\ref{eq8n} is now rewritten as:
\begin{equation}
{\partial_{{t}} {\rho}} =  {\partial^2_{{x}} {\rho}}  -  { \rho}
\label{eq8dn} 
\end{equation} 
We note for comparison that the general reaction-diffusion equation has the form \cite{xin2000front}:
\begin{equation}
{\partial_{{t}} {\rho}} =  {\partial^2_{{x}} {\rho}} +f(\rho)
\end{equation} 
where the functional $f(\rho)$ is a nonlinear term pertinent to the process under consideration (e.g. $f(\rho)=\rho (1-\rho)$ for Kolmogorov, Petrovsky, and Piskunov (KPP) nonlinearity, $f(\rho)=\rho^m (1-\rho)$ for the $m^{\text{th}}$-order Fisher nonlinearity, etc.). 
The linear approximation of the PNP model given by Eq.\;\ref{eq8dn}  has been studied by several groups including Janssen \cite{janssen2019curvature}, Janssen and Bier \cite{janssen2018transient}, Bazant et al.  \cite{bazant2004diffuse},  Singh and Kant \cite{singh2013debye, singh2014theory, birla2014theory}, and many others. It is best used for describing electrodiffusive dynamics at planar electrodes. 
However, in practice, these types of devices and systems  unavoidably exhibit in a way or another anomalies in their electrical response and frequency dispersion of their properties   due to their structural disorder, spatial heterogeneity, and wide spectrum of relaxation times. This renders the problem of describing their complex behavior restricted when using the traditional drift-diffusion model.

%\subsection{Motivation}

In particular, Eq.\;\ref{eq8dn}    considers changes in the reduced density of charge   through a control volume to be linear and memoryless, due to the fact that we only use a first-order Taylor series approximation  in space and time \cite{wheatcraft2008fractional}.  Differential equations with integer-order  differential operator are actually defined in an infinitesimally small neighborhood of the point under consideration, and therefore are a tool for describing only local media.  For the case of non-local media, the size of the control volume   must be large enough compared to the scale(s) of the heterogeneity in the medium, which makes integer-order derivatives inadequate for describing media with heterogeneity.
Furthermore, spatial heterogeneities are not necessarily static in the course of operation of the device or system, and therefore memory effects shall be taken into consideration.

  For a proper theoretical modeling of anomalous transport, one can adopt fractional calculus to include fractional time and/or spacial derivatives \cite{tarasov2022general}. This is mainly attributed to the fact that the dynamics of transport processes substantially differs from the picture of classical transport owing to   memory effects or spatial nonlocality of purely non-Markovian nature.  Fractional calculus permits to deal with  such situations via integrals and derivatives of any arbitrary real or complex order, and therefore permits to unify and extend integer-order integrals and derivatives used in classical models \cite{mainardi2022fractional, henry2010introduction, mainardi2001fundamental}. 
  Saichev and Zaslavsky
  \cite{saichev1997fractional}, Mainardi et al. \cite{mainardi2001fundamental}, and Gorenflo et al. \cite{gorenflo2000mapping} studied the  generalization of the diffusion equation with fractional derivatives with respect to time and space, in which the   first-order time derivative of the propagating quantity was replaced with a Caputo derivative and the second-order space derivative was replaced with a Riesz-Feller derivative.   Koszto{\l}owicz and Metzler 
  \cite{kosztolowicz2020diffusion} described the transport of an antibiotic in a biofilm using a time-fractional subdiffusion-absorption equation based on the Riemann- Liouville time-fractional derivative. 
Saxena, Mathai and Haubold studied extensively in  a series of papers   
\cite{saxena2004unified, saxena2014space, haubold2011further, saxena2014distributed, saxena2015computational}
 unified forms of fractional kinetic equations
and fractional reaction-diffusion equations  
in which the time derivative    is replaced by either the Caputo,    Riemann-Liouville   or a generalized fractional derivative as defined by Hilfer \cite{hilfer2000fractional},  and the space derivative   is replaced by the Riesz–Feller   derivative. Additional nonlinear terms pertinent to   reaction processes are also considered. 
Fractional reaction-diffusion equations are of specific interest in a large class of science and engineering problems for describing non-Gaussian, non-Markovian, and non-Fickian phenomena.

%\subsection{Scope of the present work}

\begin{figure}[t!]
    \centering
    \includegraphics[width=0.39\textwidth,angle=-90]{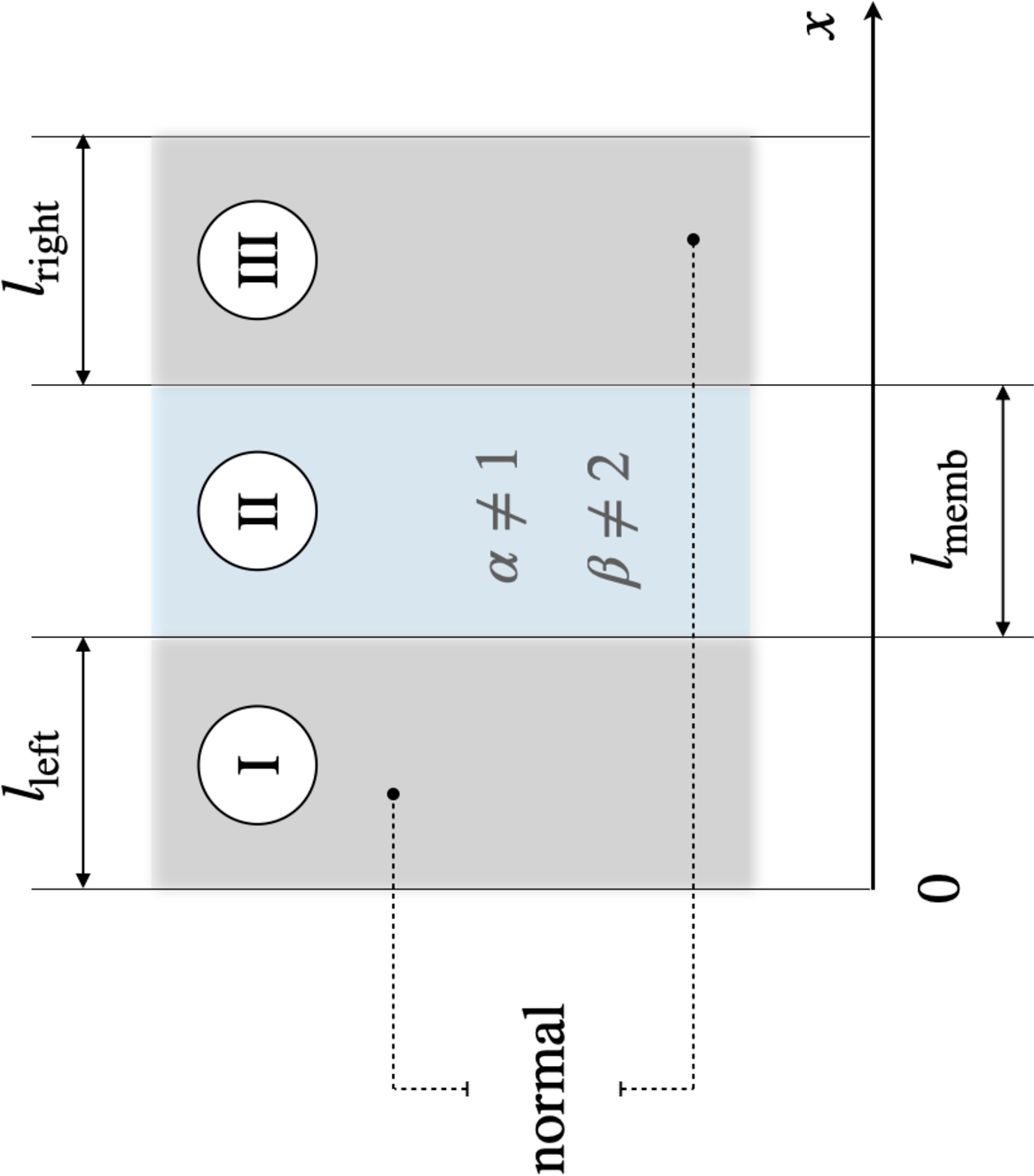}
    \caption{Nonhomogeneous membrane in which electrochemical transport is modeled with a  bi-fractional Debye-Falkenhagen equation (Eq.\;\eqref{PNP-bifrac}, $\alpha \neq 1, \beta \neq 2$ with the conditions given in \eqref{conditions}). The $x$ axis is normal to the membrane plane.}
    \label{fig:my_label}
\end{figure}

The goal of this work is to study the   bi-fractional (time and space) generalization of the (dimensionless) drift-diffusion equation of Debye and Falkenhagen (see section\;\ref{sec:model}, Eq.\;\ref{PNP-bifrac} below), and understand how do the fractional orders of differentiation affect the dynamics of the propagating quantity. 
 In section\;\ref{time-fractional} we provide the analytical solution to this equation in terms of Fox's $H$-function, followed by numerical simulations in section\;\ref{numerical-results} for different sets of values for the fractional parameters.

\section{Model}
\label{sec:model}

We consider the the  bi-fractional drift-diffusion equation in one dimension given by:
\begin{equation}
^\mathrm{c}D_t^\alpha\rho=D_{x}^{\beta,\theta}\rho-\rho
\label{PNP-bifrac}
\end{equation} 
 subjected to the boundary and initial conditions 
\begin{equation}\label{conditions}
  \rho(x=\pm\infty,t) = 0,
  \qquad
  \rho(x,t=0) = \delta(x).
\end{equation}
This model is a generalization of Eq.\;\ref{eq8dn}, and can describe for example the situation of anomalous ion diffusion through a membrane as shown in Fig.\;\ref{fig:my_label}. 
In Eq.\;\eqref{PNP-bifrac}, the operator  $^{\rm c}D^{\alpha}_t$ is the Caputo time fractional derivative of
order $\alpha$ ($0 < \alpha < 1$) replacing the first order time
derivative in Eq.\;\ref{eq8dn}, and $D^{\beta,\theta}_x$ is the Riesz-Feller space fractional derivative of order $\beta$ ($0 < \beta < 2$)
replacing the second order space derivative~\cite{mainardi2001fundamental}. The Caputo time-fractional derivative of order $\alpha$ ($m-1<\alpha<m, m\in \mathbb{N}$) of $f(t)$  is defined through the Laplace transform  ($\tilde{f}(s) = \mathcal{L}[f(t);s] = \int_0^{\infty} e^{-st} f(t) dt,\; s\in\mathbb{C}$) by:
\begin{equation}\label{Caputo-Laplace}
  \mathcal{L}\left\lbrace ^{\rm c}D^{\alpha}_t f(t) ; s  \right\rbrace = s^{\alpha} \tilde{f}(s) - \sum_{r=0}^{m-1} s^{\alpha - r - 1} f^{(r)}(0),
\end{equation}
This lead to  the integro-differential definition:
\begin{equation}\label{Caputo-fracder}
  ^{\rm c}D^{\alpha}_t  f(t) \equiv \frac{1}{\Gamma(m-\alpha)} \int\limits_0^t \frac{f^{(m)}(\tau) d\tau}{(t-\tau)^{\alpha+1-m}},
\end{equation} 
  that takes into account all past activities of the function up to the current time. 
For the case of $\alpha=m$, we have the traditional, memoryless integer-order derivative:
\begin{equation}
    ^{\rm c}D^{\alpha}_t  f(t) = \frac{d^m f(t)}{dt^m}
\end{equation} 
Whereas for a sufficiently well-behaved function $f(x)$, 
the Riesz-Feller space-fractional derivative of order $\beta$ ($0<\beta \leqslant 2$) and skewness $\theta$ ($|\theta| < \min\left\lbrace \beta, 2-\beta \right\rbrace$) is defined in terms of its Fourier transform ($ \hat{f}(k) = \mathcal{F}\{ f(x);k\} = \int_{-\infty}^{\infty} e^{ikx} f(x) dx, \; k\in\mathbb{R} $) as \cite{mainardi2001fundamental}:
\begin{equation}\label{RF-fracder}
  \mathcal{F}\left\lbrace D^{\beta,\theta}_x f(x) ; k  \right\rbrace = -|k|^\beta e^{i (\mathrm{sgn} k)\theta \pi/2} \hat{f}(k)
\end{equation}
In terms of integral representation,   the Riesz-Feller derivative can be  represented by: \cite{saxena2014distributed}:
\begin{align} 
    D^{\beta,\theta}_x f(x)  & =   \frac{\Gamma(1+\beta)}{\pi} \times \nonumber \\
    & \left\{ \sin\left[(\beta+\theta)\pi/2 \right] \int\limits_0^{\infty} \frac{f(x+\xi) - f(x)}{ \xi^{1+\beta}} d\xi\ \right. \nonumber \\
    & \left.  \sin\left[(\beta-\theta)\pi/2 \right] \int\limits_0^{\infty} \frac{f(x -\xi) - f(x)}{ \xi^{1+\beta}} d\xi   \right\}
\end{align}
For the specific case of $\theta=0$, we have the symmetric operator with respect to $x$ that can be interpreted as:
\begin{equation} 
    D^{\beta,0}_x f(x)   =  -\left[-\frac{d^2}{dx^2}\right]^{\beta/2}
\end{equation}
and Eq.\;\ref{RF-fracder} reduces to:
\begin{equation} 
  \mathcal{F}\left\lbrace D^{\beta,0}_x f(x) ; k  \right\rbrace = -|k|^\beta   \hat{f}(k)
\end{equation}

\section{Analytical solutions}
\label{time-fractional}

\subsection{Case with $0 <\alpha < 1$, $\beta=2$}
We start with the simple case of $\beta  = 2$ and skewness $\theta=0$, 
which makes Eq.~\eqref{PNP-bifrac} to reduce  to the time fractional equation of the form
\begin{equation}\label{PNP-time-frac}
  ^{\rm c}D^{\alpha}_t \rho = \partial^2_{x} \rho - \rho.
\end{equation}
Taking into account the Laplace transform of the Caputo fractional time derivative, 
  Eq.~\eqref{PNP-time-frac} in the Laplace space takes the form:
\begin{equation}\label{PNP-time-frac-L}
  s^{\alpha}\tilde{\rho}(x,s)  - s^{\alpha-1} \rho(x,0) = \partial^2_{x} \tilde{\rho}(x,s) - \tilde{\rho}(x,s).
\end{equation}
Using \eqref{conditions} and making the Fourier transform 
for both sides of Eq.~\eqref{PNP-time-frac-L}, we come to
\begin{equation}\label{PNP-time-frac-LF}
  s^{\alpha}\hat{\tilde{\rho}}(k,s)  - s^{\alpha-1} = -k^2 \hat{\tilde{\rho}}(k,s) - \hat{\tilde{\rho}}(k,s).
\end{equation}
Thus, the solution of Eq.\;\eqref{PNP-time-frac} in the Laplace-Fourier space reads,
\begin{equation}\label{PNP-time-frac-LF-sol}
  \hat{\tilde{\rho}}(k,s)  = \frac{s^{\alpha-1}}{s^\alpha + 1 + k^2}.
\end{equation}

\subsubsection{Solution in the real-Laplace space}
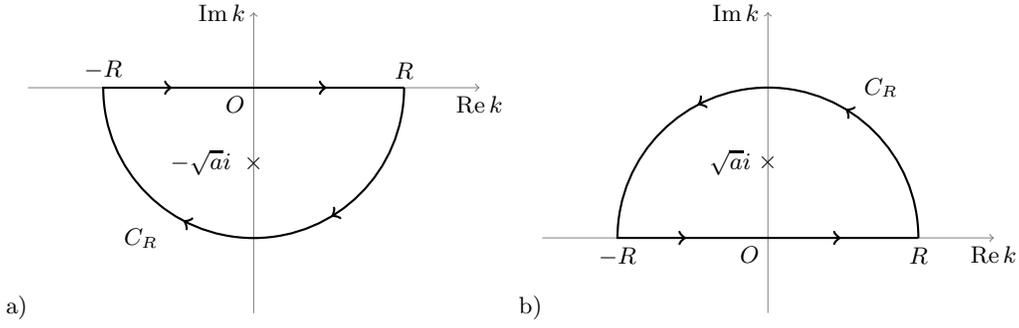
\begin{figure}[t!]
\centering
a)\begin{tikzpicture}
  [
    decoration={%
      markings,
      mark=at position 0.2 with {\arrow[line width=1pt]{>}},
      mark=at position 0.4 with {\arrow[line width=1pt]{>}},
      mark=at position 0.7 with {\arrow[line width=1pt]{>}},
      mark=at position 0.9 with {\arrow[line width=1pt]{>}},
    }
  ]
  \draw [help lines,->] (-3,0) -- (3,0) coordinate (xaxis);
  \draw [help lines,->] (0,-3) -- (0,1) coordinate (yaxis);
  \node at (0,-1) {$\times$};
  \node at (-0.7,-1) {$-\sqrt{a}i$};
  \path [draw, line width=0.8pt, postaction=decorate] (2,0) node [above] {$R$} arc (0:-180:2) node [above] {$-R$} -- (2,0);
  \node [below] at (xaxis) {$\Re k$};
  \node [left] at (yaxis) {$\Im k$};
  \node [below left] {$O$};
  \node at (-1.5,-2) {$C_{R}$};
\end{tikzpicture}
b)\begin{tikzpicture}
  [
    decoration={%
      markings,
      mark=at position 0.2 with {\arrow[line width=1pt]{>}},
      mark=at position 0.4 with {\arrow[line width=1pt]{>}},
      mark=at position 0.7 with {\arrow[line width=1pt]{>}},
      mark=at position 0.9 with {\arrow[line width=1pt]{>}},
    }
  ]
  \draw [help lines,->] (-3,0) -- (3,0) coordinate (xaxis);
  \draw [help lines,->] (0,-1) -- (0,3) coordinate (yaxis);
  \node at (0,1) {$\times$};
  \node at (-.5,1) {$\sqrt{a}i$};
  \path [draw, line width=0.8pt, postaction=decorate] (2,0) node [below] {$R$} arc (0:180:2) node [below] {$-R$} -- (2,0);
  \node [below] at (xaxis) {$\Re k$};
  \node [left] at (yaxis) {$\Im k$};
  \node [below left] {$O$};
  \node at (1.5,2) {$C_{R}$};
\end{tikzpicture}
\caption{The integration contours for (a) $x>0$  and (b)  $x<0$, and poles of the integrand on the left-hand side for Eq.\;\eqref{PNP-time-frac-L-sol-2}.}
\label{fig:contour}
\end{figure}

To get the solution in the real space, it is convenient to make the inverse Laplace and Fourier transforms with respect to $s$ and $k$, sequentially \cite{mainardi2001fundamental}. 
However, we might be interested in the solution obtained by the inverse Fourier transform with respect to $k$ and remained in the Laplace space with respect to time $t$. Formally, one can write this solution in the form
\begin{equation}\label{PNP-time-frac-L-sol-1}
  \tilde{\rho}(x,s)  = \frac{1}{2\pi} \int\limits_{-\infty}^{\infty} \frac{s^{\alpha-1}}{s^{\alpha} + 1 + k^2} e^{-ixk}\,dk.
\end{equation}
Introducing the notation $s^\alpha+1 = a$ ($\mathrm{Re}(s)>0$ and $\mathrm{Re}(a)>0$), we have 
\begin{eqnarray}
  \tilde{\rho}(x,s) & = & \frac{s^{\alpha-1}}{2\pi} \int\limits_{-\infty}^{\infty} \frac{e^{-ixk}\,dk}{a + k^2} \nonumber \\
  & = & \frac{s^{\alpha-1}}{2\pi} \int\limits_{-\infty}^{\infty} \frac{e^{-ixk}\,dk}{(k - \sqrt{a}i)(k + \sqrt{a}i)}. \label{PNP-time-frac-L-sol-2}
\end{eqnarray}
The integrand in Eq.~\eqref{PNP-time-frac-L-sol-2}
is analytic everywhere except for the isolated
singularities $k = \pm \sqrt{a} i$, where it has simple poles. 
For $x>0$, using the residue theorem, we have 
\begin{eqnarray}
  \lim_{R \to \infty} & & \oint_{C_R} \frac{e^{-i k x} \, dk}
  {(k - \sqrt{a}i)(k + \sqrt{a}i)}  \nonumber \\
  &=& -2\pi i \, \mathrm{res}_{k = -\sqrt{a}i} \left[ \frac{ e^{-i k x}}{(k - \sqrt{a}i)(k + \sqrt{a}i)}\right], \label{residue-theorem}
\end{eqnarray}
where the contour $C_R$ is shown in Fig.\ref{fig:contour}a. 
As $R \to \infty$, 
the integral over the arc of the circle tends to zero, because the integrand
\begin{equation*}
 \frac{e^{-i kx}}
  {(k + \sqrt{a}i)(k - \sqrt{a}i)}
  =  \frac{ e^{-i x\Re k} e^{x\Im k}}{(k + \sqrt{a}i)(k - \sqrt{a}i)},
  \quad
  \Im k < 0
\end{equation*}
vanishes exponentially for $x>0$. Therefore, 
\begin{equation}\label{lim-R}
   \lim_{R \to \infty} \oint_{C_R} \frac{e^{-i k x} \, dk}
  {(k - \sqrt{a}i)(k + \sqrt{a}i)} 
  = \int\limits_{-\infty}^{\infty}  \frac{e^{-i kx} \, dk}{(k + \sqrt{a}i)(k - \sqrt{a}i)}.
\end{equation}
Calculating the residue, we obtain
\begin{equation}
  \mathrm{res}_{k = -\sqrt{a}i} 
  \left[ \frac{e^{-i kx}}{(k + \sqrt{a}i)(k - \sqrt{a}i)} \right]
  = -\frac{e^{-\sqrt{a}x}}{2\sqrt{a}i}.
\end{equation}
Substituting the latter and \eqref{lim-R} in Eq.~\eqref{residue-theorem}, we obtain
\begin{equation}\label{countour-res}
  \int\limits_{-\infty}^{\infty}  \frac{e^{-i kx} \, dk}{(k + \sqrt{a}i)(k - \sqrt{a}i)}
  = \frac{\pi e^{-\sqrt{a}x}}{\sqrt{a}},
  \quad x >0.
\end{equation}
Thus, Eq.~\eqref{PNP-time-frac-L-sol-2} takes the form 
\begin{equation}\label{PNP-time-frac-L-sol-3}
  \tilde{\rho}(x,s)  = \frac{s^{\alpha-1}e^{-\sqrt{a}x}}{2\sqrt{a}},
  \quad
  x > 0.
\end{equation}
Similarly, for $x<0$ we consider the contour $C_R$ is shown in Fig.\ref{fig:contour}b. 
The result for Eq.~\eqref{PNP-time-frac-L-sol-2} in this case reads,
\begin{equation}\label{PNP-time-frac-L-sol-4}
  \tilde{\rho}(x,s)  = \frac{s^{\alpha-1}e^{\sqrt{a}x}}{2\sqrt{a}},
  \quad
  x < 0.
\end{equation}
Thus, combining Eqs.~\eqref{PNP-time-frac-L-sol-3} and \eqref{PNP-time-frac-L-sol-4} together, we come to
\begin{equation}\label{PNP-time-frac-L-sol-fin}
  \tilde{\rho}(x,s)  = \frac{s^{\alpha-1}e^{-\sqrt{a}|x|}}{2\sqrt{a}}.
\end{equation}
Finally, using $a = s^{\alpha} + 1$, we obtain 
\begin{equation}\label{PNP-time-frac-L-sol-fin1}
  \tilde{\rho}(x,s)  = \frac{s^{\alpha-1}}{2(s^{\alpha} + 1)^{1/2}} \exp\left[- |x| (s^{\alpha}+1)^{1/2} \right].
\end{equation}

We should note that for the time-fractional diffusion equation 
\begin{equation}\label{PNP-time-frac-simple}
  ^{\rm c}D^{\alpha}_t \rho = \partial^2_{x} \rho,
\end{equation}
the solution in the Laplace-Fourier space reads,
\begin{equation}\label{PNP-time-frac-simple-LF-sol}
  \hat{\tilde{\rho}}(k,s)  = \frac{s^{\alpha-1}}{s^\alpha + k^2}.
\end{equation}
Thus, from Eq.~\eqref{PNP-time-frac-L-sol-fin} with $a = s^\alpha$, one can get the solution in agreement with \cite{MK00},
\begin{equation}\label{PNP-time-frac-simple-L-sol-fin}
  \tilde{\rho}(x,s)  = \frac12 s^{\alpha/2-1} \exp\left[ -|x| s^{\alpha/2} \right].
\end{equation}

\subsubsection{Solution in the Fourier-time space}

Unfortunately,  the inverse Laplace transform of Eq.~\eqref{PNP-time-frac-L-sol-fin1} is problematic.
However, we can invert the Laplace transform from Eq.~\eqref{PNP-time-frac-LF-sol} following  Langlands\;\cite{Lan06}.
We rewrite Eq.~\eqref{PNP-time-frac-LF-sol} as 
\begin{equation}\label{PNP-time-frac-LF-sol-1}
  \hat{\tilde{\rho}}(k,s)  = \frac{s^{\alpha-1}}{s^\alpha + 1 + k^2} = \frac{s^{\alpha-1}}{s^\alpha + k^2} \frac{1}{1 + \frac{1}{s^{\alpha} + k^2}}.
\end{equation}
Now by expanding the second fraction we have
\begin{equation}\label{PNP-time-frac-LF-sol-2}
  \hat{\tilde{\rho}}(k,s)  = \frac{s^{\alpha-1}}{s^\alpha + k^2} \sum_{r=0}^\infty \frac{(-1)^r}{(s^\alpha + k^2)^r}
  = \sum_{r=0}^\infty \frac{(-1)^r s^{\alpha-1} }{(s^\alpha + k^2)^{r+1}}.
\end{equation}
From \cite{Pod99} we have the following Laplace transform
\begin{equation}\label{Laplace-ML}
  \mathcal{L}\left\lbrace t^{\alpha r + \beta - 1} E^{(r)}_{\alpha,\beta}(-at^\alpha) ; s \right\rbrace
  = \frac{r! s^{\alpha-\beta}}{(s^\alpha + a)^{r+1}},
\end{equation}
where 
\begin{equation}\label{Mittag-Leffler}
  E_{\alpha,\beta}(z) = \sum_{k=0}^\infty \frac{z^k}{\Gamma(\alpha k + \beta)}
\end{equation}
is the Mittag-Leffler function. Thus, using Eq.~\eqref{Laplace-ML} with $a=k^2$ and $\beta = 1$, 
we can invert the Laplace transform in \eqref{PNP-time-frac-LF-sol-2} to get
\begin{equation}\label{PNP-time-frac-F-sol}
  \hat{\rho}(k,t)  
  = \sum_{r=0}^\infty \frac{(-1)^r t^{\alpha r} }{r!} E_{\alpha}^{(r)} (-k^2 t^\alpha).
\end{equation}
The derivatives of the Mittag-Leffler function can be expressed in terms of the $H$-function (see Appendix\;\ref{H-function}) ~\cite{Lan06,MSH10},
\begin{equation}\label{ML-H}
  E^{(r)}_{\alpha,\beta}(-z)  = H^{1,1}_{1,2}\left[ z \bigr| 
  {\scriptsize \begin{matrix} (-r, 1) \\
  (0,1), (-\alpha r, \alpha) 
  \end{matrix}}   \right], 
\end{equation}
knowing that the generalized Mittag-Leffler function in terms of the Mellin-Barnes integral representation is given by \cite{saxena2004unified}:
\begin{equation}
    E_{\alpha,\beta}^{\gamma}(z) = \frac{1}{\Gamma(\gamma)}\frac{1}{2\pi i} \int_{\Omega} \frac{\Gamma(-\xi) \Gamma(\gamma + \xi) (-z)^{\xi} d\xi}{\Gamma( \alpha \xi + \beta)}
\end{equation}
and thus:
\begin{equation}\label{Mittag-Leffler3}
    E_{\alpha,\beta}^{\gamma}(z) = H^{1,1}_{1,2}\left[ z \bigr| 
  {\scriptsize \begin{matrix} (1-\gamma, 1) \\
  (0,1), (1-\beta, \alpha) 
  \end{matrix}}   \right].
\end{equation}
The two-parameter Mittag-Leffler (Eq.\;\ref{Mittag-Leffler}) is obtained by setting $\gamma=1$ in Eq.\;\ref{Mittag-Leffler3}. 
Now one can then rewrite Eq.~\eqref{PNP-time-frac-F-sol} in the form 
\begin{equation}\label{PNP-time-frac-F-sol-1}
  \hat{\rho}(k,t)  
  = \sum_{r=0}^\infty \frac{(-1)^r t^{\alpha r} }{r!} H^{1,1}_{1,2}\left[ k^2 t^\alpha \bigr| 
  {\scriptsize \begin{matrix} (-r, 1) \\
  (0,1), (-\alpha r, \alpha) 
  \end{matrix}} \right].
\end{equation}

\subsubsection{Solution in the real-time space}
Now we invert the Fourier transform in Eq.~\eqref{PNP-time-frac-F-sol-1}. To do this, we note that $\hat{\rho}(k,t)$
is an even function of $k$. For an even function $\hat f(k) = \hat f(-k)$, 
the  Fourier transform reduces to the Fourier cosine transform,
\begin{eqnarray}
  f(x) = \frac{1}{2\pi} \int\limits_{-\infty}^{\infty} \hat f(k) e^{-ikx} \, dk  
  %& = & \frac{1}{2\pi} \left[ \int_{0}^{\infty} \hat f(k) e^{-ikx} \, dk + \int_{-\infty}^{0} \hat f(k) e^{-ikx} \, dk  \right] \\
  %& = &  \frac{1}{2\pi} \left[ \int_{0}^{\infty} \hat f(k) e^{-ikx} \, dk + \int_{\infty}^{0} \hat f(-k) e^{ikx} \, d(-k)  \right] \\
  %& = &    \frac{1}{2\pi} \left[ \int_{0}^{\infty} \hat f(k) e^{-ikx} \, dk + \int_{0}^{\infty} \hat f(k) e^{ikx} \, dk  \right] \\
   =  \frac{1}{\pi} \int\limits_{0}^{\infty} \hat f(k) \cos(kx) \, dk.
\end{eqnarray}
The inverse Fourier cosine transform can be calculated using the following relation for the cosine transform of the $H$-function~\cite{SMH04}
\begin{eqnarray}
  \int\limits_0^{\infty} k^{\rho-1} \cos(kx) H^{m,n}_{p,q} \left[ a k^\mu \bigr|
  {\scriptsize \begin{matrix} (a_p, A_p) \\
  (b_q,B_q)
  \end{matrix}}
   \right] \, dk \nonumber \\
   = \frac{\pi}{x^\rho} H^{n+1,m}_{q+1,p+2} \left[ \frac{x^\mu}{a} \bigr|
  {\scriptsize \begin{matrix} (1-b_q, B_q),(\frac12+\frac{\rho}{2},\frac{\mu}{2}) \\
  (\rho,\mu), (1-a_p,A_p),(\frac12+\frac{\rho}{2},\frac{\mu}{2})
  \end{matrix}}
   \right]. \label{cosine-H}
\end{eqnarray}
Using the latter with $\rho=1$, $a = t^\alpha$, $\mu=2$, and $m,n,p,q$, $(a_p,A_p)$ and $(b_q,B_q)$
coefficients defined in Eq.~\eqref{PNP-time-frac-F-sol-1}, one can invert the Fourier transform in Eq.~\eqref{PNP-time-frac-F-sol-1}
to obtain
\begin{eqnarray}
  \rho(x,t)  
  & = & \frac{1}{\pi} \sum_{r=0}^\infty \frac{(-1)^r t^{\alpha r} }{r!}  \int\limits_0^{\infty} \cos(kx) H^{1,1}_{1,2}\left[ k^2 t^\alpha \bigr| 
  {\scriptsize \begin{matrix} (-r, 1) \\
  (0,1), (-\alpha r, \alpha) 
  \end{matrix}} \right] \, dk \nonumber \\
  & = &  \frac{1}{|x|}  \sum_{r=0}^\infty \frac{(-1)^r t^{\alpha r} }{r!} H^{2,1}_{3,3} \left[ \frac{x^2}{t^\alpha} \bigr| 
  {\scriptsize \begin{matrix} (1, 1), (1+\alpha r, \alpha), (1,1) \\
  (1,2), (1+r, 1), (1,1) 
  \end{matrix}} \right].  
  \label{PNP-time-frac-sol}
\end{eqnarray}
Next, using the following reduction formula~\cite{MSH10}
\begin{eqnarray}
  H^{m,n}_{p,q}\left[ z \bigr| 
  {\scriptsize \begin{matrix} (a_1, A_1),\ldots,(a_p,A_p) \\
  (b_1,B_1),\ldots,(b_{q-1},B_{q-1}), (a_1,A_1) 
  \end{matrix}} \right] \nonumber \\
  = H^{m,n-1}_{p-1,q-1}\left[ z \bigr| 
  {\scriptsize \begin{matrix} (a_2, A_2),\ldots,(a_p,A_p) \\
  (b_1,B_1),\ldots,(b_{q-1},B_{q-1})
  \end{matrix}} \right],\label{H-reduction}
\end{eqnarray}
we can simplify Eq.~\eqref{PNP-time-frac-sol} to 
\begin{equation}\label{PNP-time-frac-sol-1}
  \rho(x,t)  
  = \frac{1}{|x|}  \sum_{r=0}^\infty \frac{(-1)^r t^{\alpha r} }{r!} 
  H^{2,0}_{2,2} \left[ \frac{x^2}{t^\alpha} \bigr| 
  {\scriptsize \begin{matrix} (1+\alpha r, \alpha), (1,1) \\
  (1,2), (1+r, 1)
  \end{matrix}} \right]. 
\end{equation}
Finally, using the property of the $H$-function~\cite{MSH10},
\begin{equation}\label{H-power}
  z^\sigma H^{m,n}_{p,q}\left[ z \bigr| 
  {\scriptsize \begin{matrix} (a_p,A_p) \\
  (b_q,B_q)
  \end{matrix}} \right]
  = H^{m,n}_{p,q}\left[ z \bigr| 
  {\scriptsize \begin{matrix} (a_p+\sigma A_p,A_p) \\
  (b_q+\sigma B_q,B_q)
  \end{matrix}} \right],
  \;
  \sigma \in \mathbb{C},
\end{equation}
with $\sigma = -1/2$, we come to 
\begin{equation}\label{PNP-time-frac-sol-fin}
  \rho(x,t)  
  = \sum_{r=0}^\infty \frac{(-1)^r t^{\alpha (r-\frac12)} }{r!} 
  H^{2,0}_{2,2} \left[ \frac{x^2}{t^\alpha} \bigr| 
  {\scriptsize \begin{matrix} (1+\alpha [r-\frac12], \alpha), (\frac12,1) \\
  (0,2), (r+\frac12, 1)
  \end{matrix}} \right]. 
\end{equation}

Together, Eqs.~\eqref{PNP-time-frac-LF-sol}, \eqref{PNP-time-frac-L-sol-fin}, \eqref{PNP-time-frac-F-sol-1} and \eqref{PNP-time-frac-sol-fin} provide the solution to the time fractional equation \eqref{PNP-time-frac} in four different spaces 
with respect to the density arguments, namely $x \leftrightarrow k$, and $t \leftrightarrow s$.

%\section{Solution to the bi-fractional time-space drift-diffusion equation}

\subsection{Case with $0 <\alpha < 1$, $0 <\beta < 2$}

The solution to the bi-fractional drift-diffusion Eq.~\eqref{PNP-bifrac} with $0 <\alpha < 1$, $0 <\beta < 2$, $\theta=0$  in real-time space can be obtained similarly to the time-fractional equation~\eqref{PNP-time-frac}. 
The Laplace-Fourier transformations of Eq.~\eqref{PNP-bifrac} with the conditions  given in \eqref{conditions}  is:
\begin{equation}\label{PNP-bi-frac-LF-sol}
  \hat{\tilde{\rho}}(k,s)  = \frac{s^{\alpha-1}}{s^\alpha + 1 + k^\beta}.
\end{equation}
The result for $\rho(x,t)$ is found to be:
\begin{equation}\label{PNP-bifrac-sol}
  \rho(x,t)  
  = \frac{1}{|x|}\sum_{r=0}^\infty \frac{(-1)^r t^{\alpha r}}{r!} 
  H^{2,1}_{3,3} \left[ \frac{|x|^\beta}{t^\alpha} \bigr| 
  {\scriptsize \begin{matrix} (1,1), (1+\alpha r, \alpha), (1,\frac{\beta}{2}) \\
  (1,\beta), (r+1, 1), (1,\frac{\beta}{2})
  \end{matrix}} \right]. 
\end{equation}
Using Eq.~\eqref{H-power} with $\sigma = -1/\beta$, one can rewrite \eqref{PNP-bifrac-sol} as 
\begin{eqnarray}
  \rho(x,t)  
  & = & \sum_{r=0}^\infty \frac{(-1)^r t^{\alpha (r - \frac{1}{\beta})}}{r!}
  \nonumber\\
  & \times & 
  H^{2,1}_{3,3} \left[ \frac{|x|^\beta}{t^\alpha} \bigr| 
  {\scriptsize \begin{matrix} (1-\frac{1}{\beta},1), (1+\alpha [r - \frac{1}{\beta}], \alpha), (\frac12, \frac{\beta}{2}) \\
  (0,\beta), (r+1-\frac{1}{\beta}, 1), (\frac12,\frac{\beta}{2})
  \end{matrix}} \right].\label{PNP-bifrac-sol-fin} 
\end{eqnarray}

\section{Numerical results}
\label{numerical-results}

\begin{figure*}[htbp]
\begin{center}
\includegraphics[height=1.7in]{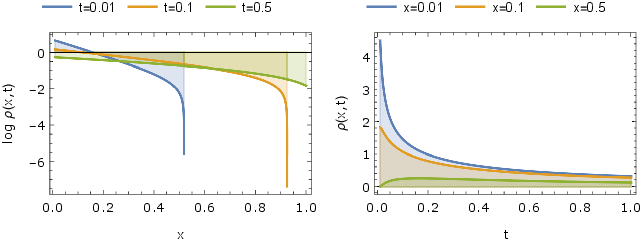}
\includegraphics[height=1.6in]{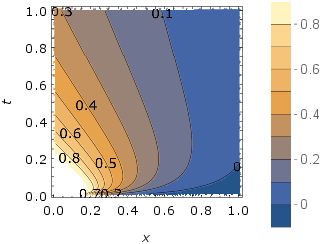}
\caption{Plots of $\rho(x,t)$ given by Eq.\;\ref{PNP-bifrac-sol-fin} with $\alpha=1.0$, $\beta=2.0$ as a function of (a) $x$ for $t=0.01$, $0.1$, $0.5$, (b) $t$ for $x=0.01$, $0.1$, $0.5$ and (c) $x$ and $t$ (contour plot).}
\label{fig3}
\end{center}
\end{figure*}

We calculate the obtained solutions for $\rho(x,t)$ governed by Eq.\;\eqref{PNP-bifrac} with the boundary and initial conditions given by\;\ref{conditions} for the four cases of (i) normal eletrodiffusion ($\alpha=1$, $\beta=2$), 
 (ii) time-fractional eletrodiffusion ($0<\alpha<1$, $\beta=2$), 
 (iii) space-fractional eletrodiffusion ($\alpha=1$, $0<\beta<2$) and (iv) bi-fractional electrodiffusion ($0<\alpha<1$, $0<\beta<2$) as given by  Eq.\;\ref{PNP-bifrac-sol-fin}. We fixed the upper limit of the summation to five terms, which is deemed sufficient to represent well enough the overall behavior of the variable $\rho(x,t)$. 
The Fox $H$-function can be calculated numerically using a simple rectangular approximation of the  integrals~\cite{AEM16}.
The function $\rho(x,t)$ is calculated for $x\in[-1,-\delta) \cup (\delta,1]$ and $t\in(\varepsilon,0.25]$,
where $\delta>0$ and $\varepsilon>0$ are utilized to cut small locality around $x=0$, $t=0$, where the Fox $H$-function and $\rho(x,t)$ 
are not defined. 
We remind again that $\rho(x,t)$ described by Eq.\;\eqref{PNP-bifrac} is a generalization of the integer-order Debye-Falkenhagen approximation (Eq.\;\ref{eq8dn}), whose validity is limited to the regime of small applied potentials. 

%Fig.~\ref{fig:timefrac} shows the solution \eqref{PNP-time-frac-sol-fin} of the time-fractional equation with $\alpha=0.7$.
%Fig.~\ref{fig:bifrac} shows the solution \eqref{PNP-bifrac-sol-fin} of the bi-fractional equation with $\alpha=0.7$ and $\beta = 1.7$.
%Fig.~\ref{fig:bifrac1} the solution \eqref{PNP-bifrac-sol-fin} of the bi-fractional equation with $\alpha=0.5$ and $\beta = 1.3$.

First we consider the known integer-order case of $\alpha=1$, $\beta=2$ (i.e. Eq.\;\ref{eq8dn}). 
It is clear that at the limit $\alpha\to 1$  we obtain from Eq.\;\ref{PNP-time-frac-sol-fin} the following expression for $\rho(x,t)$:
\begin{eqnarray}
\rho(x,t) &=&
\frac{e^{-\sqrt{\frac{x^2}{t}}}}{2 \sqrt{t}} \nonumber \\
 &+& (-\sqrt{t} +\frac{t^{3/2}}{2} -  \ldots)
  H^{1,0}_{1,1} \left[ \frac{x^2}{t} \bigr| 
  {\scriptsize \begin{matrix} (\frac{1}{2},1) \\
  (0,2)
  \end{matrix}} \right]
  \label{a1b2}
  \end{eqnarray}
  The same can be found from Eq.\;\ref{PNP-bifrac-sol-fin} for  $\alpha\to 1$, $\beta \to 2$. We recognize that the first term in Eq.\;\ref{a1b2} corresponds to the fundamental solution of the standard Fick's diffusion equation $\partial_t \rho = \partial_x^2 \rho$. Solutions to the integer-order case of Debye-Falkenhagen equation for different conditions  has been previously provided mainly via numerical simulations and approximations (e.g. by using  Pad\'{e} approximation) \cite{bazant2004diffuse,  feicht2016discharging, stout2017diffuse}, but here    by using tools from fractional calculus we give an analytical expression as an infinite series of the Fox $H$-function.    
  Plots of $\rho(x,t)$   for this case as a function of $x$ ($0.01<x<1$) for the different values of   $t=0.01$, $0.1$, $0.5$ (in log-linear scale), and as a function of $t$ ($0.01<t<1$) for the different values of $x=0.01$, $0.1$, $0.5$ (in linear-linear scale) are shows in Figs.\;\ref{fig3}(a) and\;\;\ref{fig3}(b) respectively. Figs.\;\ref{fig3}(c) is the contour  plot of   $\rho(x,t)$ depicting its spatiotemporal dynamics.  
 The solution depicting concentrations is always positive. It  is an even function of $x$ and decays to zero for large values of $|x|$. It also decays to zero for large values of $t$.
  
  \begin{figure*}[htbp]
\begin{center}
\includegraphics[height=1.7in]{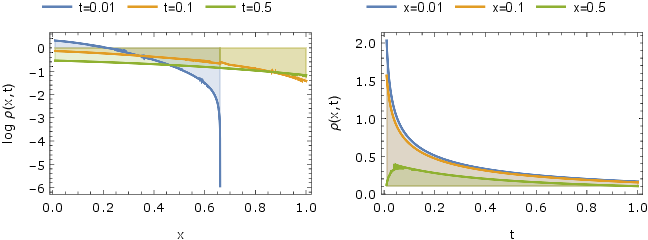}
\includegraphics[height=1.6in]{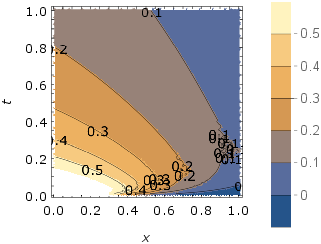}
\caption{Plots of $\rho(x,t)$ given by Eq.\;\ref{PNP-bifrac-sol-fin} with for $\alpha=0.8$, $\beta=2.0$ as a function of (a) $x$ for $t=0.01$, $0.1$, $0.5$, (b) $t$ for $x=0.01$, $0.1$, $0.5$ and (c) $x$ and $t$ (contour plot).}
\label{fig4}
\end{center}
\end{figure*}

\begin{figure*}[htbp]
\begin{center}
\includegraphics[height=1.7in]{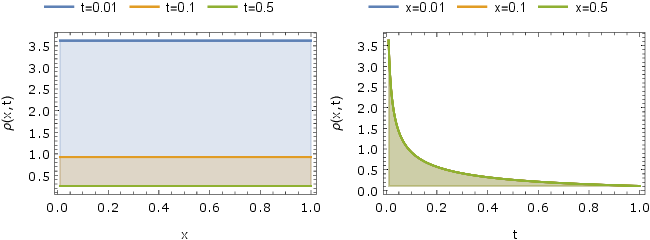}
\includegraphics[height=1.6in]{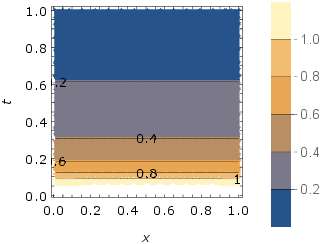}
\caption{Plots of $\rho(x,t)$ given by Eq.\;\ref{PNP-bifrac-sol-fin} with for $\alpha=1.0$, $\beta=1.8$ as a function of (a) $x$ for $t=0.01$, $0.1$, $0.5$, (b) $t$ for $x=0.01$, $0.1$, $0.5$ and (c) $x$ and $t$ (contour plot).}
\label{fig5}
\end{center}
\end{figure*}

\begin{figure*}[htbp]
\begin{center}
\includegraphics[height=1.7in]{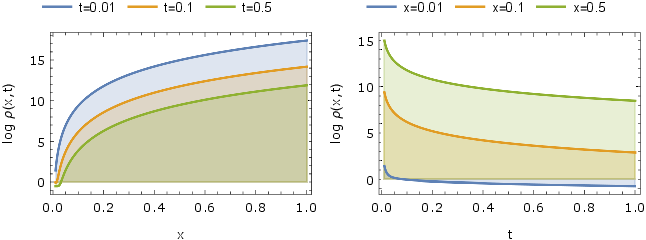}
\includegraphics[height=1.6in]{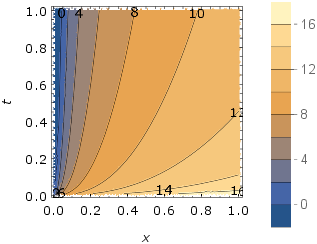}
\caption{Plots of $\log(\rho(x,t))$ given by Eq.\;\ref{PNP-bifrac-sol-fin} with for $\alpha=0.8$, $\beta=1.8$ as a function of (a) $x$ for $t=0.01$, $0.1$, $0.5$, (b) $t$ for $x=0.01$, $0.1$, $0.5$ and (c) $x$ and $t$ (contour plot).}
\label{fig6}
\end{center}
\end{figure*}

For the time-fractional anomalous case of $0 <\alpha < 1$, $\beta=2$, we verify that  Eq.\;\ref{PNP-bifrac-sol-fin} reduces to Eq.\;\ref{PNP-time-frac-sol-fin}. Similar to the previous case, plots of $\rho(x,t)$ as a function of $x$, as a function of $t$, and as a function of both $x$ and $t$ for $\alpha =0.8$, $\beta=2$ are shown in Fig.\;\ref{fig4}.

For the space-fractional anomalous case of $\alpha = 1$, $0<\beta<2$,   Eq.\;\ref{PNP-bifrac-sol-fin} simplifies to: 
\begin{eqnarray}
\rho(x,t) = (t^{-1/\beta } &-&t^{1-1/\beta } + \frac{1}{2}  t^{2-1/\beta } -\ldots) \nonumber \\
&\times &
H_{2,2}^{1,1}
\left[ \frac{x^{\beta}}{t} \bigr| 
  {\scriptsize \begin{matrix} (1-\frac{1}{\beta},1), (\frac{1}{2}, \frac{\beta}{2}) \\
  (0,\beta), (\frac{1}{2}, \frac{\beta}{2}) 
  \end{matrix}} \right]
\end{eqnarray}
  which is plotted in Fig.\;\ref{fig5}  for the case of $\alpha = 1$, $\beta=1.8$.
  
  Finally, in Fig.\;\ref{fig6} we show the variation of  $\log \rho(x,t)$   vs. both variables $x$ and vs. $t$ for the general case of  two fractional parameters,   $\alpha = 0.8$ and $\beta=1.8$. The propagating quantity $\rho(x,t)$ tends to accelerate as $x$ and $t$ increase, and thus the representation in log scale.

%\begin{figure}[ht!]
%\centering
%\includegraphics[width=1.0\linewidth]{density_time_frac.png}
%\caption{Solution $\rho(x,t)$ of the time-fractional equation \eqref{PNP-time-frac} calculated using analytic expression~\eqref{PNP-time-frac-sol-fin} for $\alpha=0.7$.}
%\label{fig:timefrac}
%\end{figure} 
%
%\begin{figure}[ht!]
%\centering
%\includegraphics[width=1.0\linewidth]{density_bi_frac.png}
%\caption{Solution $\rho(x,t)$ of the bi-fractional equation \eqref{PNP-bifrac} 
%calculated using analytic expression~\eqref{PNP-bifrac-sol-fin} for $\alpha=0.7$ and $\beta = 1.7$.}
%\label{fig:bifrac}
%\end{figure} 
%
%\begin{figure}[ht!]
%\centering
%\includegraphics[width=1.0\linewidth]{density_bi_frac1.png}
%\caption{Solution $\rho(x,t)$ of the bi-fractional equation \eqref{PNP-bifrac} 
%calculated using analytic expression~\eqref{PNP-bifrac-sol-fin} for $\alpha=0.5$ and $\beta = 1.3$.}
%\label{fig:bifrac1}
%\end{figure} 

%\FloatBarrier
%
%{\color{red}
%For impedance theory, see~\cite{BE05,LEB09}. In our model with fractional derivatives for both space and time, I did not understand how to adopt this approach straightforwardly. 
%}
%
%
%\FloatBarrier

\section{Conclusion}

The electrochemical modeling of electrified porous structures in contact with an electrolyte is   quite challenging. The traditional mathematical tools are based on integer-order differential equations, which are more suited for homogeneous systems with planar geometries. 
When complex structures and coupled phenomena are involved, it is often required to further complement the existing models by additional approximations and assumptions which renders the problem even more difficult to solve. The theoretical and numerical results presented in this work show the possibilities that come with the use of both time and space bi-fractional-order derivatives for the case   of the Debye-Falkenhagen equation, which is a simple and idealized model for  electrodiffusion at low applied voltages. Eq.\;\ref{PNP-bifrac-sol-fin}, with its extra two degrees of freedom, $\alpha$ and $\beta$, compared to the integer-order model (Eq.\;\ref{a1b2}) is capable of deforming the spatiotemporal dynamics of the propagating quantity $\rho(x,t)$ in ways to account for subdiffusive and superdiffusive transports. 
While the physical interpretations of the fractional parameters remains unclear and need further studies, the mathematical solutions to this general problem can provide useful insights in anomalous electrodiffusion in heterogeneous media such as  membranes, protein channels  and electrochemical devices.

\appendix
\section{Fox's $H$-function\label{H-function}}

The Fox's $H$-function \cite{fox1961g} is defined by means  of a Mellin-Barnes  type  integral in the following manner \cite{mathai2009h}:
\begin{align}
H^{m,n}_{p,q}(z) &= H^{m,n}_{p,q}\left[ z|^{(a_p,A_p)}_{(b_q,B_q)} \right] \nonumber \\
&=H^{m,n}_{p,q}\left[ z|^{(a_1,A_1),\ldots,(a_p,A_p)}_{(b_1,B_1),\ldots, (b_q,B_q)} \right] \nonumber \\
&=\frac{1}{2\pi i} \int_L h(s) z^{-s} ds
\end{align}
 with $h(s)$  given by the  ratio of products of Gamma functions:
 \begin{equation}
h(s) = \frac{
\prod_{j=1}^m \Gamma(b_j + B_j s) \; 
\prod_{j=1}^n \Gamma(1-a_j - A_j s) 
}
{
\prod_{j={n+1}}^p \Gamma(a_j + A_j s)\;
\prod_{j={m+1}}^q \Gamma(1-b_j - B_j s) 
}
\label{eq:h}
\end{equation}
$m,n,p,q$ are  integers satisfying ($0 \leqslant n \leqslant p$, $1 \leqslant m \leqslant q$), 
$z\neq 0$, and $z^{-s}=\exp \left[ -s (\ln|z|+ i \arg z) \right] $,  $A_i, B_j \in \mathbb{R}_+$, $a_i, b_j \in \mathbb{R}$ or $\mathbb{C}$ with $(i=1,2,\ldots,p)$, $(j=1,2,\ldots,q)$. 
The contour of integration $L$ is a suitable contour separating the poles $-(b_j+\nu)/B_j$, ($j=1,\ldots,m$; $\nu=0, 1, 2, \ldots$),  of the gamma functions $\Gamma(b_j+ B_j s)$ from the poles $(1-a_{\lambda} +k)/A_{\lambda}$, ($\lambda=1,\ldots,n$; $k=0, 1, 2, \ldots$) of the gamma functions  $\Gamma (1-a_{\lambda} - A_{\lambda} s)$, that is $A_{\lambda} (b_j+ \nu) \neq B_j (a_{\lambda - k - 1})$. 
 An empty product in\;\ref{eq:h}, if it occurs, is taken to be one. 
 The $H$-function contains a vast number of elementary and special functions as special cases. 
Detailed and comprehensive accounts of the matter are available in Mathai,  Saxena, and   Haubold \cite{mathai2009h}, Mathai and  Saxena \cite{mathai1978h}, and Kilbas and Saigo \cite{saigo2004h}

%\bibliography{bifrac_notes.bib}

%merlin.mbs apsrev4-1.bst 2010-07-25 4.21a (PWD, AO, DPC) hacked
%Control: key (0)
%Control: author (8) initials jnrlst
%Control: editor formatted (1) identically to author
%Control: production of article title (-1) disabled
%Control: page (0) single
%Control: year (1) truncated
%Control: production of eprint (0) enabled
%

\end{document}